\documentclass[aps,pra,twocolumn,showpacs,superscriptaddress,letterpaper]{revtex4-1}

\usepackage{graphicx}
\usepackage{subfigure}
\usepackage{color}
\usepackage{physymb}
\usepackage{enumitem}
\usepackage{mdframed}
\usepackage{framed}
\usepackage{epstopdf}

\definecolor{red}{rgb}{1,0,0}
\definecolor{orange}{rgb}{1,0.5,0}
\definecolor{green}{rgb}{0.13,0.55,0.13}
\definecolor{purple}{rgb}{0.5,0,1}

\begin{document}

\title{Development and Uses of Upper-division Conceptual Assessments}


\author{Bethany R. Wilcox}
\affiliation{Department of Physics, University of Colorado, 390 UCB, Boulder, CO 80309}

\author{Marcos D. Caballero}
\affiliation{Department of Physics and Astronomy \& CREATE for STEM Institute, Michigan State University, East Lansing, MI 48824}

\author{Charles Baily}
\affiliation{School of Physics and Astronomy, University of St.\ Andrews, St.\ Andrews, Fife KY16 9SS Scotland, UK}

\author{Homeyra Sadaghiani}
\affiliation{Department of Physics and Astronomy, California State Polytechnic University, Pomona, CA 91768}

\author{Stephanie V. Chasteen}
\affiliation{Department of Physics, University of Colorado, 390 UCB, Boulder, CO 80309}

\author{Qing X. Ryan}
\affiliation{Department of Physics, University of Colorado, 390 UCB, Boulder, CO 80309}

\author{Steven J. Pollock}
\affiliation{Department of Physics, University of Colorado, 390 UCB, Boulder, CO 80309}

\begin{abstract}
The use of validated conceptual assessments alongside more standard course exams has become standard practice for the introductory courses in many physics departments.  These assessments provide a more standard measure of certain learning goals, allowing for comparisons of student learning across instructors, semesters, and institutions.  Researchers at the University of Colorado Boulder have developed several similar assessments designed to target the more advanced physics content of upper-division classical mechanics, electrostatics, quantum mechanics, and electrodynamics.  Here, we synthesize the existing research on our upper-division assessments and discuss some of the barriers and challenges associated with developing, validating, and implementing these assessments as well as some of the strategies we have used to overcome these barriers.  
\end{abstract}

\maketitle

\section{Introduction}

Research-based conceptual assessments represent one of the most commonly adopted tools to come out of the Physics Education Research (PER) community in the last several decades.  At the introductory level, the Force Concept Inventory \cite{hestenes1992fci} is arguably the most well known of these assessments; however, many other instruments, spanning multiple topical areas, have been developed (see Ref.\ \cite{hestenes1992fci, thornton1998fmce, ding2006bema} for examples and Ref.\ \cite{beichnerAssessment} for a more comprehensive list).  Historically, these assessments have had a number of significant impacts on physics education at the introductory level.  For example, they provide a measure of some aspects of student learning that are often not captured by standard exams.  They also represent a standardized and validated tool for evaluating the effectiveness of classroom strategies across instructors, institutions, and time. By providing a measure of student learning across courses and pedagogies, these assessments can support both pedagogical and institutional changes that enhance student learning \cite{hake1998ie, chasteen2012transforming}.  

Fewer conceptual assessments have been developed to target upper-division physics content, in part because conceptual assessment at the advanced undergraduate level presents some uniques challenges.  For example, advanced physics content requires students to manipulate sophisticated mathematical tools and techniques.  This increased emphasis on mathematics makes it more difficult, and perhaps less desirable, to create assessments that focus only on students' conceptual understanding.  Additionally, the increased complexity of the physics content makes it challenging to construct clear, level-appropriate questions that can be answered within a reasonably short time frame.  The relatively small body of existing research on students' difficulties also makes it more difficult to create questions that specifically target areas where students struggle.  Various logistical constraints to the development of standardized assessments also become more of a barrier at the upper-division level than the introductory level.  For example, less consistency in content coverage between different instructors and institutions makes it more difficult to create a single instrument that matches the learning goals of a majority of courses/instructors.  Additionally, small class sizes at the upper-division level hinder efforts to collect enough early-implementation data to achieve sufficient statistical power to ensure the validity and reliability of a new instrument.  

Despite these challenges, several conceptual assessments have been developed for the upper-division level, targeting a range of content areas that include (but are not limited to): sophomore classical mechanics \cite{caballero2013ccmi}, junior electricity and magnetism \cite{notaros2002emci, chasteen2012cue, baily2012materials}, quantum mechanics \cite{zhu2012qms, cataloglu2002qmvi, goldhaber2009qmat}, and several engineering assessments targeting thermodynamics \cite{midkiff2001tci, streveler2011ttci, evans2003CIs} and waves \cite{thoads1999wave}.  Additionally, several assessments developed for the introductory and sophomore level have been used productively as pre/post tests at the upper-division level \cite{singh2006scge, mckagan2010qmcs}.  Note that, here, we are using the term `conceptual assessment' broadly and include in this category assessments that target aspects of mathematical thinking (rather than procedural mathematics) and strategic processes and practices (e.g., identifying the correct solution method). 

\begin{table*}
\begin{tabular}{  p{5.3cm} p{0.2cm} p{1.4cm}  p{.7cm} p{0.2cm} p{6.6cm}  p{0.2cm} p{1.8cm} }
   \hline
   \hline
   {\bf Assessment} & & {\bf Question Format} & {\bf \# of Q's} & & {\bf Level and Content} & & {\bf Standard Text} \\
   \hline
   Colorado Classical Mechanics/Math Methods Instrument ({\bf CCMI}) \cite{caballero2013ccmi} & & FR & 11* & & Sophomore mechanics up to (but not including) the Lagrangian and Hamiltonian formulations & & Taylor \cite{taylor2005classmech} \hspace{7mm}Ch. 1-5 ** \\
   Colorado Upper-division Electrostatics Diagnostic ({\bf CUE}) \cite{chasteen2012transforming} & & FR $^\dagger$ & 17 & & Junior electrostatics with minimal magnetostatics & & Griffiths \cite{griffiths1999em} \hspace{9mm}Ch. 1-5  \\
   Colorado UppeR-division ElectrodyNamics Test ({\bf CURrENT}) \cite{baily2012materials, ryan2014current} & & FR & 6 & & Junior electrodynamics up to (but not including) relativistic electrodynamics & & Griffiths \cite{griffiths1999em} \hspace{9mm}Ch. 7-9\\
   Quantum Mechanics Assessment Tool ({\bf QMAT}) \cite{goldhaber2009qmat} & & FR $^\dagger$ & 14 & & Junior quantum mechanics focusing on solutions to the time-independent Schr\"{o}dinger equation & & Griffiths \cite{griffiths1995qm} \hspace{9mm}Ch. 1-4  \\
   \hline
   Electromagnetics Concept Inventory ({\bf EMCI}) \cite{notaros2002emci} & & MC & 23*** & & Electrodynamics for junior-level engineers including both fields and waves & & NA \\
   Symmetry and Gauss's Law Conceptual Evaluation ({\bf SGLCE}) \cite{singh2006scge} & & MC & 25 & & Conceptual understanding of symmetry and Gauss's Law at the introductory level & & NA \\
   Quantum Mechanics Survey ({\bf QMS}) \cite{zhu2012qms} & & MC & 31 & & Junior quantum mechanics in one spacial dimension & & NA \\
   Quantum Mechanics Visualization Instrument ({\bf QMVI}) \cite{cataloglu2002qmvi} & & MC & 25 & & Introductory through graduate quantum with an emphasis on visualization & & NA \\
   Quantum Mechanics Conceptual Survey ({\bf QMCS}) \cite{mckagan2010qmcs} & & MC & 12 & & Quantum mechanics as appropriate for sophomore-level modern physics & & NA \\
   \hline
   \hline
\end{tabular}
\caption{Brief overview of specific upper-division conceptual assessments. The top section includes CU's four named upper-division assessments, and the bottom section includes similar information for several of the major alternative instruments that have been developed for or used at the upper division level. Each assessment is classified as either multiple-choice (MC) if the final numerical score comes from only multiple-choice or multiple-response questions, or free-response (FR) otherwise.  \\ $^\dagger$ Multiple-choice/multiple-response version of these assessments have been created and will be discussed in greater detail in Sec.\ \ref{sec:scoring}.  \\ * Only 9 questions on the CCMI contribute to the overall numerical score. The remaining 2 questions target the use of specific mathematical tools (Fourier Series and separation of variables) and are outside the scope of most classical mechanics courses.  \\ ** Coverage includes Newton's Universal Law of Gravitation following Ch. 5 of Thornton and Marion \cite{marion2003classical}. \\ *** The EMCI is split into two, 23 question versions: one targeting fields and one targeting waves.   }\label{tab:instruments}
\end{table*}

In this paper, we will focus on four assessments created at the University of Colorado Boulder (CU) as examples of the development and uses of upper-division conceptual assessments.  The goals of this paper are to: (1) present an overview of CU’s upper-division conceptual assessments including motivation, approaches, development, and current status of each instrument (Sec.\ \ref{sec:CUassessments}) while highlighting similarities and differences between our approach and that of others; (2) summarize examples of outcomes from each assessment (Sec.\ \ref{sec:uses}); and (3) discuss implementation of these assessments in the classroom including barriers and possible solutions (Sec.\ \ref{sec:implementation}).  We will not be presenting new findings that, for example, compare learning outcomes or unpack student difficulties, but rather we will present an overview of conceptual assessment for upper-division courses.

\section{\label{sec:CUassessments}Upper-division Conceptual Assessments at CU}

Over the past 8 years, the PER group at CU has developed four conceptual assessments for the upper-division level: the Colorado Classical Mechanics/Mathematical Methods Instrument (CCMI), the Colorado Upper-division Electrostatics Diagnostic (CUE), the Colorado UppeR-division ElectrodyNamics Test (CURrENT), and the Quantum Mechanics Assessment Tool (QMAT).  The development of these instruments was motivated, in part, by a need for a research-based and validated measure of the success of our course transformation efforts \cite{chasteen2012transforming, pepper2011faculty, goldhaber2009qmat, baily2012materials} with respect to learning goals developed for each course.  These learning goals were developed in conjunction with a broad cross-section of CU physics faculty as part of the Science Education Initiative's model for course transformation \cite{chasteen2014sei}.  The goals represent a consensus of what these faculty want students to be able to do after completing our upper-division courses \cite{pepper2011faculty}.  Although they were developed locally, these goals are not specific to the courses taught at CU, and feedback from external faculty suggest that these learning goals are valued and relevant more broadly in the U.S.\ physics community (see Ref.\ \cite{sei} for the full list of learning goals).  In particular, we were interested in designing our assessments to target specific learning goals that were not typically assessed by traditional exams (e.g., conceptual understanding).  

Each assessment was designed to target topics in one of the canonical core upper-division content areas (e.g., electrostatics, quantum mechanics, etc.); however, they are not designed to provide a comprehensive assessment of all material.  The goal was instead to focus on a smaller subset of the material in order to provide a litmus test for student achievement with respect to our learning goals. A brief overview of each of our four named assessments is given in Table \ref{tab:instruments} along with information on several other assessment instruments that target the same core content areas.  Comparisons of the development and validation of these instruments will be discussed later.

\subsection{\label{content}Content Coverage}

\subsubsection{Electricity and Magnetism}

For the first half of junior electricity and magnetism, three potential assessment instruments are (see Table \ref{tab:instruments}): the CUE, the Symmetry and Gauss's Law Conceptual Evaluation (SGLCE), and the Electromagnetics Concept Inventory - Fields (EMCI - Fields).  The SGLCE was designed to assess introductory physics students' ideas about symmetry and Gauss's Law \cite{singh2006scge}, but preliminary testing with upper-division undergraduates and graduate students suggest that this instrument is challenging even for more advanced students.  However, as an introductory assessment, the SGLCE does not include any of the more advanced electrostatics topics (e.g., solutions to Laplace's Equation).  The EMCI - Fields was designed to target electrostatics, magnetostatics, and time-varying electromagnetic fields for junior engineering courses \cite{notaros2002emci}.  The content coverage of the CUE is similar to that of the EMCI, but the CUE does not include time-dependence, as this is typically not included in a first semester electricity and magnetism course in physics departments \cite{chasteen2012cue}.  

For the second half of junior electricity and magnetism, two assessments are available: the CURrENT, and the Electromagnetics Concept inventory - Waves (EMCI - Waves).  The EMCI - Waves focuses exclusively on the propagation and generation of electromagnetic waves, with a strong emphasis on engineering applications (e.g., waveguides, transmission lines, etc.) \cite{notaros2002emci}.  The CURrENT picks up where the CUE leaves off with time-variation, electromagnetic waves, and Maxwell's Equations \cite{ryan2014current}.  Neither of these instruments includes relativistic electrodynamics.  Note that Notaros \emph{et al.} \cite{notaros2002emci} have also crafted a 25 question combined EMCI that would be appropriate for a single semester electromagnetism course and covers a sampling of topics from both the Waves and Fields versions of the assessment.  

\subsubsection{Quantum Mechanics}

A relatively large number of assessments have been developed for quantum mechanics including (see Table \ref{tab:instruments}): the QMAT, the Quantum Mechanics Survey (QMS), the Quantum Mechanics Visualization Instrument (QMVI), and the Quantum Mechanics Conceptual Survey (QMCS).  Of these, only the QMAT and QMS were specifically developed for the upper-division level, and both target measurement, solutions to the Schr\"{o}dinger equation in one dimension, and time-evolution from a wavefunctions-first perspective \cite{goldhaber2009qmat, zhu2012qms}.  The QMVI was designed to provide a longitudinal measure of students' understanding from introductory up through graduate quantum mechanics \cite{cataloglu2002qmvi} with a specific emphasis on visualization.  The longitudinal focus of the QMVI means that it includes a number of topics not typically covered until graduate quantum mechanics \cite{mckagan2010qmcs}.  Lastly, the QMCS was developed to target sophomore-level, introductory quantum mechanics (i.e., modern physics).  While the developers suggest that the QMCS may be particularly valuable as a pre-post measure in more advanced courses, they also note that many faculty see the QMCS as too basic for the upper-division level \cite{mckagan2010qmcs}.  

\subsubsection{Classical Mechanics and Thermodynamics}

There are a number of assessments that target mechanics at the introductory level (e.g., \cite{hestenes1992fci, thornton1998fmce}; however, we are aware of only one published instrument for mechanics beyond the introductory level (i.e., sophomore-level classical mechanics), the CCMI.  The CCMI was developed to target mechanics up to (but not including) the Lagrangian and Hamiltonian formulations, as well as gravitation \cite{caballero2013ccmi}.  Additionally, while there are several engineering focused thermodynamics inventories available \cite{midkiff2001tci, streveler2011ttci, evans2003CIs}, we are not aware of any published, physics-centric thermodynamic instruments.

\subsection{\label{sec:development}Development}

\begin{figure}
\includegraphics[height=3in, width=1.8in]{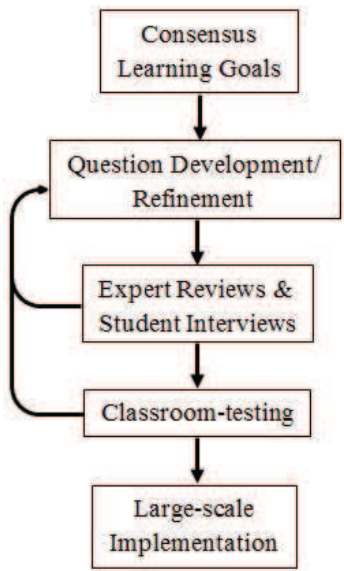}
\caption{A schematic of the design process used to develop CU's upper-division conceptual assessments.}\label{fig:dCycle}
\end{figure}

The development of each of CU’s four upper-division assessments followed the same basic process (See Fig.\ \ref{fig:dCycle}).  The first draft of each assessment was generated in faculty working groups facilitated by PER specialists/postdocs, who then further developed and refined the instruments.  Initial question development was focused on addressing course-scale learning goals \cite{sei} articulated through collaborative discussions with CU physics faculty who had taught each course \cite{pepper2011faculty}.  These learning goals guided all stages of the assessment development including content coverage and format.  For example, these consensus learning goals motivated one of the more unique aspects of CU's assessments: the free-response format.  Because our learning goals emphasized students' ability to synthesize, generate, and justify their responses, we determined that an open-ended format would be more valued by faculty.  Early drafts were also informed by known student difficulties identified through informal observations (e.g., in-lecture discussions or in small group work) and (when available) research on students' understanding.  

When developing upper-division assessments, one of the key challenges is creating level-appropriate questions that can be answered within a reasonably short time frame.  For example, we quickly found that students are easily slowed by complicated calculations or tasks that required remembering exact formulae.  To avoid excessive time spent on calculations, we used a number of different strategies such as asking students only to provide and justify a solution method rather than having them actually work through a problem.  Another strategy that proved effective was asking students only to provide the sign of a certain quantity or whether it was zero or non-zero, rather than asking them to determine the value of that quantity.  

After a preliminary draft was completed, each assessment was reviewed by multiple experts in either physics content or assessment.  Expert reviews establish the content validity of the assessments by ensuring that: (1) the physics content was accurate and clearly expressed, and (2) this content was valued by experts and perceived as appropriate for the upper-division level.  The assessments were revised and refined based on this feedback.  For example, early drafts of the assessments were often too long, and expert feedback was vital to shortening the instruments by identifying and eliminating questions that were least reflective of the goals of the course.   

Revised drafts were then given to a small number (5-15) of volunteer undergraduates in an interview setting.  Student interviews establish the face validity of the assessments by confirming that students were interpreting the prompts and figures as we intended.  Interviews were conducted with individual students in a think-aloud format where interviewees were asked to articulate their reasoning as they worked through the problems on the assessment.  When necessary, the assessments were modified to ensure that a student's score reflected their knowledge of the physics content and not their understanding (or lack thereof) of the question.

Following expert reviews and student interviews, the assessments were tested as in-class post-tests in at least one semester of the associated course.  Student scores during the classroom testing phase were analyzed to identify questions that were too difficult or too easy, or that did not discriminate between high and low achieving students (see Sec.\ \ref{sec:validation}).  These items were either removed or modified, and the new version was re-tested with students in interviews and in-class implementations.  Classroom testing is also critical to ensuring that the majority of students can complete the assessment within a 50 minute period.  For example, early tests of the CURrENT indicated that the instrument was too long and several questions/subparts were removed as a result.  The final version of each assessment was a result of iterative refinement based on expert reviews, student interviews, and student performance in classroom tests (see Fig.\ \ref{fig:dCycle}).  

The available literature on other upper-division assessments (see Table \ref{tab:instruments}) suggests they were developed using a similar iterative design cycle.  One notable difference is the central role that our explicitly-articulated learning goals played in the design process.  These meta-level goals guided us towards developing questions that not only targeted content knowledge, but also assessed reasoning and problem solving skills (e.g., Fig.\ \ref{fig:complexRubric}).  Alternatively, literature on the development of other assessments focuses on content coverage, typically determined through textbook reviews and faculty surveys.  Specific questions are often developed to target known student difficulties; however, alignment of the questions and overall instrument with explicitly-articulated learning goals is not typically discussed for other assessments.

\subsection{\label{sec:scoring}Scoring}

\begin{figure}
    \begin{minipage}{.95\linewidth}
    \flushleft \emph{Give a brief outline of the EASIEST method you would use to solve the problem.  \\ {\bf DO NOT SOLVE the problem, we just want to know: \\ \hspace{2mm}(1) The general strategy (half credit) and \\ \hspace{2mm}(2) Why you chose that method (half credit)}}
    \end{minipage}
    \begin{minipage}{.68\linewidth}
        \vspace{2mm}\flushleft {\bf Q7}. \emph{A solid non-conducting sphere, centered on the origin, with a non-uniform charge density that depends on the distance from the origin, $\rho(r)=\rho_o e^{-r^2/a^2}$ where $a$ is a constant.  \\Find E (or V) at point P.}
    \end{minipage}
    \begin{minipage}{.25\linewidth}
        \includegraphics[height=20mm, width=20mm]{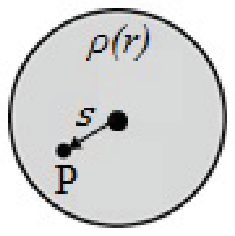}
    \end{minipage}
    \begin{minipage}{1\linewidth}
        \vspace{2mm}
         \begin{tabular}{ p{2cm}  p{5.9cm} }
            {\bf Q7 Rubric} &  \\
            \hline
            \hline
            Correct & Correct answer is Gauss's Law \\
            Answer & +1 point for saying direct integration \\
            (3 pts) & \\
            \hline
            Explanation (2 pts)& Full credit requires some explanation of why (not just how) Gauss's Law is used.  This would include some mention of the Gaussian surface used or the symmetry (such as charge distribution depends only on ``r'', or E field is radial).\\
             &{\bf+1} point if the correct Gaussian surface is drawn \\
             &{\bf+0-1} point for explaining how to solve by Gauss's Law \\
             & If answer ``direct integration'' must give explanation of how they would solve this integral. {\bf 0.5} for a poor explanation of how they would go about it (e.g., writing down Coulomb's Law) \\
            \hline
            \hline
        \end{tabular}
    \end{minipage}
\caption{An example question from the CUE along with the associated scoring rubric.  This style of complex rubric was used for the CUE and QMAT and graders must undergo specific training to use these rubrics to produce reliable scores.}\label{fig:complexRubric}
\end{figure}

During the classroom testing phase of developing these assessments, it is necessary to develop scoring rubrics.  For open-ended conceptual assessments this process is particularly challenging as any rubric must allow multiple graders to produce valid and reliable scores.  We have utilized two distinct styles of grading rubrics with our conceptual assessments (discussed below).  

The CUE and QMAT were the first of the CU assessments to be developed, and both assessments are characterized by fairly open-ended questions (e.g., Fig.\ \ref{fig:complexRubric}).  These open-ended questions have the potential to elicit a large variety of student responses: however, creating clear, reliable rubrics for such questions requires complex and nuanced grading schemes that reflect subtle differences in students responses.  To create these rubrics, common student ideas on each question were identified from student responses during classroom testing, and the developers agreed on scores for these common answers.  A detailed grading scheme describing specific point allocations for a variety of student responses was developed to reflect these consensus scores.  An example of this style of rubric is given in Fig.\ \ref{fig:complexRubric}.    

Early tests of inter-rater reliability for the complex grading rubrics for the CUE and QMAT (see Fig.\ \ref{fig:complexRubric}) showed that some amount of training was necessary for new graders to produce consistent scores.  For the case of the CUE, training involves a new grader independently scoring 10-15 example CUE exams and comparing their results to established scores.  Conservative measures of the inter-rater reliability of the final version of the CUE scoring rubric indicated a substantial degree of inter-rater reliability when defining agreement as no more than a 10\% difference between scores on each question by different graders \cite{chasteen2012cue}.  While effective at ensuring reliable scores between different graders, the training process is time consuming (roughly 5 hrs) and thus represents a significant barrier to faculty adoption.  

When work on the CUE began, we were not aware of any existing examples of such a complex validated rubric for a free-response conceptual assessment that could be used reliably by independent graders.  While the work with the CUE ultimately demonstrated that creating this type of rubric was possible, its development required dedicated work by a PER post-doc and multiple iterations of reliability testing.  Efforts to create an equally reliable grading rubric for the QMAT were less successful in part because the QMAT intentionally provides multiple opportunities for even more broadly open-ended responses (e.g. ``describe what happens to the real and imaginary parts as time goes by, with words and pictures'', ``list important qualitative features'', ``give an example of a state which...'').  While these questions elicited rich and informative student responses, they made creation of unambiguous grading criteria more difficult. When the post-doc responsible for the QMAT left before reliability testing could be completed, further development was put on hold. Recent efforts have shifted the nature of the instrument by redesigning it in a multiple choice format (discussed briefly below).  

\begin{figure}
    \begin{minipage}{.95\linewidth}
        \vspace{-2mm}\flushleft{\bf Q3}) \emph{Below is a plot of the potential energy, in Joules, of a particle free to move on a 2-d plane. Values of the potential energy are given for darkened contours. }

\vspace*{-12pt}
\begin{center}
\includegraphics[height=50mm, width=50mm]{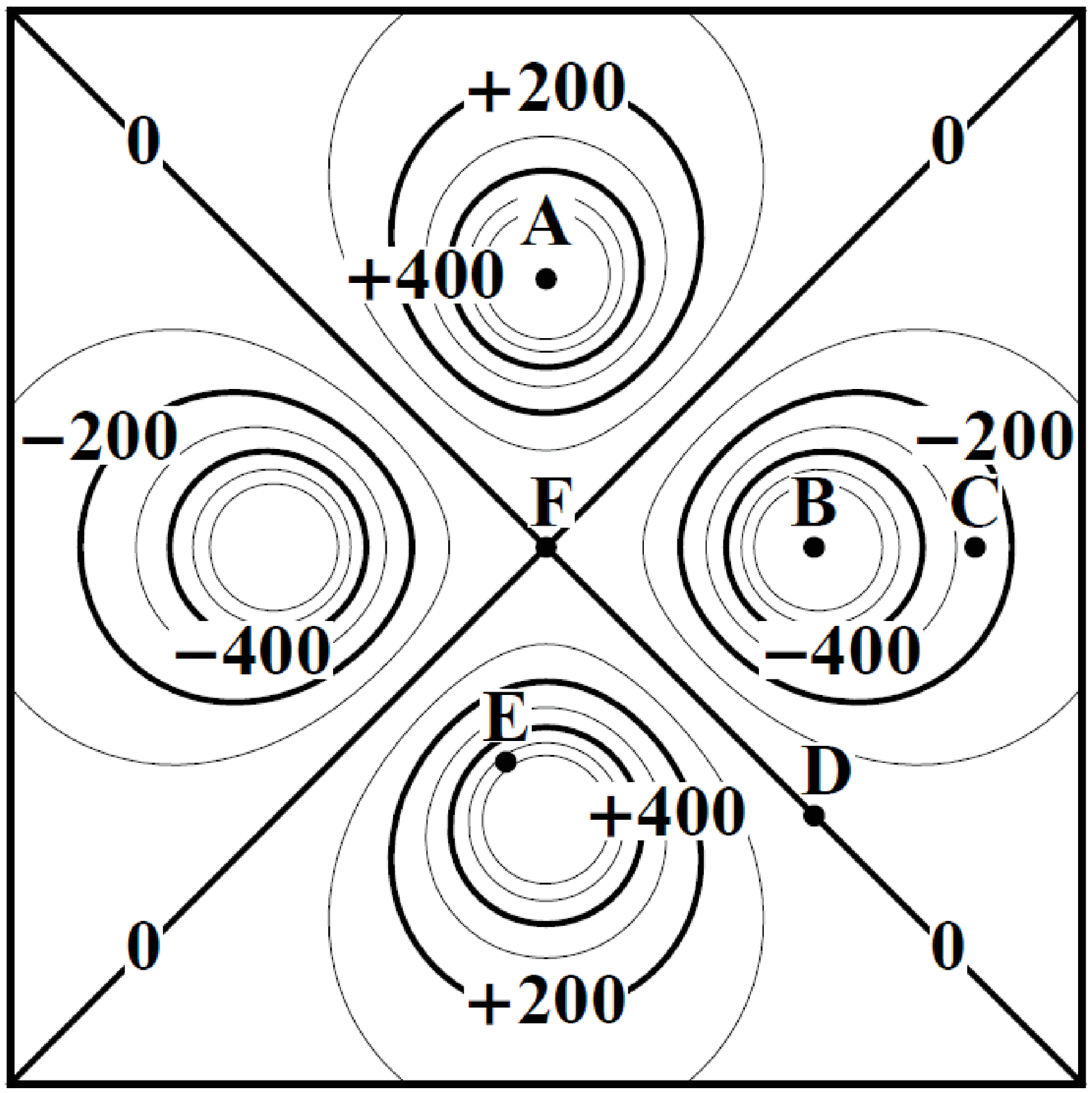}
\end{center}
\vspace*{-12pt}

\emph{(a) For which of these points (A--F) is the particle in stable equilibrium? \\
(b) Please explain how you decided on the above answer. \\
(c) Rank the magnitude of the gradient at the above points, from largest to smallest. (If
some points have gradients with equal magnitude, please make that clear in your
answer.)} 
    \end{minipage}
    \begin{minipage}{1\linewidth}
       \vspace{2mm}\flushleft {\bf Part A} (1 pt)
        \begin{tabular}{ p{1.9cm}  p{1.6cm}  p{4.7cm} }
            \hline
            \hline
            Full credit, 1 & Correct &Only (B) \\
            No credit, 0 & Incorrect & Any other responses \\
            \hline
            \hline
        \end{tabular}
        \flushleft {\bf Part B} (1 pt)
        \begin{tabular}{ p{1.9cm}  p{1.6cm}  p{4.7cm} }
            \hline
            \hline
            Full credit, 1 & Correct & Reasonable answer to \emph{correct response only}, for example: \\
            & & - Lowest potential energy, valley/well analogy, stable against small pushes \\
            No credit, 0 & Incorrect & Any other responses \\
            \hline
            \hline
        \end{tabular}
        \flushleft {\bf Part C} (1 pt)
        \begin{tabular}{ p{1.9cm}  p{1.6cm}  p{4.7cm} }
            \hline
            \hline
            Full credit, 1 & Correct & Either E$>$C$>$D$>$A=B=F or E$>$C$>$D=A=B=F \\
            & & {\bf (it is unclear if there's a small non-zero gradient at D)} \\
            & & Give credit even if signs are missing (e.g., E,C,D,(A,B,F)) \\
            Half credit, & Incomplete & Correct but missing 1 location \\
            0.5 & answer &(e.g., E$>$C$>$A=B=F) \\
            No credit, 0 & Incorrect & Any other responses \\
            \hline
            \hline
        \end{tabular}
    \end{minipage}
\caption{An example question from the CCMI along with the associated scoring rubric.  This style of simple rubric was used for the CCMI and CURrENT and graders do not need to undergo specific training to produce reliable scores.} \label{fig:simpleRubric}
\end{figure}

Informed by the difficulties encountered in the development and use of the complex scoring rubrics for the CUE and QMAT, questions on the CCMI and CURrENT were intentionally designed to support less ambiguity in scoring \cite{caballero2013ccmi}.  This necessitated a shift away from the more open-ended questions that characterize the CUE and QMAT.  Instead, the CCMI and CURrENT questions (e.g., Fig.\ \ref{fig:simpleRubric}) were designed to elicit a more constrained set of student responses while still capturing aspects of student reasoning.   The scoring rubrics for the CCMI and CURrENT emphasize identifying correct elements and have fewer opportunities for partial credit than in the rubrics for the CUE and QMAT.  An example of this style of rubric is given in Fig.\ \ref{fig:simpleRubric}.  

These rubrics were developed and validated using the same process as the rubrics for the CUE and QMAT; however, the `all or nothing' focus makes the CCMI and CURrENT rubrics considerably simpler \cite{ryan2014current, doughty2014ccmiDifficulties}.  Typically, these grading schemes do not award points based on intermediate steps or for partially correct or incomplete responses.  Minimal training is required to achieve a high degree of inter-rater reliability using these rubrics \cite{ryan2014current, doughty2014ccmiDifficulties}.  One trade-off of the simpler rubrics is that they do not typically include as many examples of common incorrect responses that can help an instructor recognize or anticipate common student difficulties.  To counteract this, we have begun creating a second `difficulties' rubric for the CCMI \cite{doughty2014ccmiDifficulties}.  This rubric is not designed to provide numerical scores, but instead to describe common student difficulties with each of the questions and presents examples of what these difficulties look like.  

Motivated in part by pressure from external institutions to simplify the scoring process further, we have recently begun exploring the viability of various multiple-choice versions of these assessments.  To date we have developed multiple-response and multiple-choice versions of the CUE and QMAT.  Distractors for both of these instruments were constructed from common responses to the free-response versions.  The new version of the CUE utilizes a novel `select all' format, which we are calling coupled multiple-response (CMR) \cite{wilcox2014cmr}.  The new version of the QMAT, now called the Quantum Mechanics Conceptual Assessment (QMCA) \cite{sadaghiani2013qmca, sadaghiani2014qmca}, is being developed by author H.S. and uses a standard multiple-choice format.   

\subsection{\label{sec:validation}Validation}

Once reliable scoring rubrics were developed and sufficient data collected during classroom testing, we generated test statistics that establish the validity and reliability of our new assessment instruments.  Two common perspectives on test development are Classical Test Theory (CTT) \cite{engelhardt2009ctt} and Item Response Theory (IRT) \cite{ding2009mcanalysis}.  The majority of conceptual assessments in physics, both at the introductory and upper-division level have been validated using CTT, while only a small number have been developed or analyzed using IRT \cite{ding2014bemaVrasch, aslanides2013rciVirt, marshall2009taksVirt, planinic2010fciVrasch}.  One significant drawback of CTT is that all test statistics are population dependent.  As a consequence, there is no guarantee that test statistics calculated for one student population (e.g., physics students at a community college) will hold for another population (e.g., physics students at a university).  For additional discussion of the limitations of CTT, see Ref.\ \cite{wallace2010irt}.  

IRT was specifically designed to address the shortcomings of CTT and all item and student parameters are independent of both population and test form \cite{ding2009mcanalysis}.  However, there are several significant drawbacks to IRT as a potential tool to develop upper-division physics assessments.  Even the most simple dichotomous IRT models require large N ($>$100) to produce reliable estimates of item and student parameters \cite{deayala2009polyIRT, ding2009mcanalysis}.  This number increases for more complex models that, for example, include item discrimination parameters, or for instruments with polytomous scoring \cite{deayala2009polyIRT}.  The small class sizes typical of upper-division physics would necessitate classroom testing at multiple institutions, possibly over multiple semesters, to collect this volume of data.  Due in large part to the logistical barriers to IRT, the development and validation of our upper-division assessments was guided by CTT. 

CTT posits several characteristics of a high quality assessment and a number of test statistics that quantify how well an instrument matches these characteristics.  For polytomously scored assessments, these statistics include \cite{engelhardt2009ctt}: item difficulty as measured by the average score on each individual item, item discrimination as measured by Pearson Correlation Coefficients of item scores with the rest of the test, internal consistency as measured by Cronbach's Alpha \cite{cortina1993ca}, and whole test discrimination as measured by Ferguson's Delta \cite{ding2006bema}.  For dichotomously scored assessments, several of the test statistics used are slightly different (see Ref.\ \cite{engelhardt2009ctt} for an overview).  

\begin{table}
\begin{tabular}{ l  r r r }
    \hline
    \hline
    Assessment \hspace{4mm}& Years of & Status of & \hspace{2mm}Validation \\
    & \hspace{1mm} Active Work & \hspace{2mm} Instrument & Statistics? \\
    \hline
    CCMI & 3 & Near final & Yes \cite{caballero2013ccmi} \\
    CUE & 5 & Finalized & Yes \cite{chasteen2012cue} \\
    CURrENT & 3 & Near final & Yes \cite{ryan2014current} \\
    QMAT & 2 & Archived & No \\
    \hline
    CMR CUE & 2 & Near final & Yes \cite{wilcox2014cmr} \\
    QMCA & 2 & Near final & Yes \cite{sadaghiani2014qmca} \\
    \hline
    \hline
\end{tabular}
\caption{Status of the development and validation of CU's upper-division assessments.  The bottom two assessments are newly developed multiple-response and multiple-choice versions of the CUE and QMAT.  }\label{tab:aStatus}
\end{table}

Because work on each of the CU upper-division assessments began at different times, each is currently at a slightly different stage of development and validation (see Table \ref{tab:aStatus}).  The CUE is the oldest of the assessments and has had nearly 7 years of continuous work including development and data collection.  The CUE is available in its final form with full validation statistics \cite{chasteen2012cue}.  Development of the CCMI and CURrENT began roughly 3 years ago and are both in the final stages of classroom testing.  These instruments will only undergo minor revision before final publication.  All preliminary test statistics indicate that both assessments are valid and reliable \cite{caballero2013ccmi, ryan2014current}.  Development of the QMAT began shortly after the CUE ($\sim$7 years ago) and continued for roughly 2 years.   However, development of the QMAT was put on hold before classroom testing was completed, and validation statistics were never published for this instrument.  Work on the multiple-response CUE and QMCA began roughly 2 years ago, and both instruments are in the final stages of validation \cite{wilcox2014cmr, sadaghiani2014qmca}. 

As an example of test validation using CTT, we summarize here the validation statistics for those assessments listed in Table \ref{tab:aStatus} (other than the QMAT); published statistics for other upper-division assessments (Table \ref{tab:instruments}) also tend to fall within the same ranges.  Overall student performance across courses and institutions is between 45-55\% for all of our instruments.  These averages, while low compared to traditional course exams, allow for considerable discrimination between high and low performers.  Consistent with this, all of our instruments have Ferguson's Delta values of $\delta \ge$ 0.97, where anything above 0.9 is considered acceptable \cite{ding2006bema}.  Additionally, all items on these assessments have item-test correlation coefficients above the standard cuttof (r $\ge$ 0.2 \cite{ding2006bema}), demonstrating a satisfactory degree of item discrimination.  In terms of internal consistency, all of our assessments have Cronbach's Alpha values of $\alpha \ge$ 0.75 with the exception of the CURrENT which has $\alpha$ = 0.69 when treating numbered questions as items (N=6) and $\alpha$ = 0.72 when treating numbered sub-parts as items (N=15).  While $\alpha$ for the CURrENT is closer to the standard threshold ($\alpha \ge$ 0.7), it has also been shown that having fewer test items tends to drive Cronbach's Alpha downward \cite{cortina1993ca}.  The CURrENT, with only 6 questions or 15 sub-parts, is most susceptible to this tendency.  Thus, all of CU's upper-division assessments provide valid and reliable measures of student learning for the tested population of students.

\section{\label{sec:uses}Uses of CU's Upper-Division Assessments}

Once developed, student performance on these assessments can be used for a variety of different purposes by researchers, administrators, and individual instructors.  

\subsection{\label{sec:learning}As a measure of student learning}

As with conceptual assessments at the introductory level, the most common motivation for using upper-division assessments is as a standardized measure of student performance that can be compared across time, courses, instructors, institutions, and pedagogies.  Indeed, one of the primary motivators for the development of our instruments was a need to assess the effectiveness of our upper-division course transformation efforts relative to other instructional practices \cite{chasteen2014sei}.  

\begin{figure}
    \includegraphics[height=2.1in, width=3.1in]{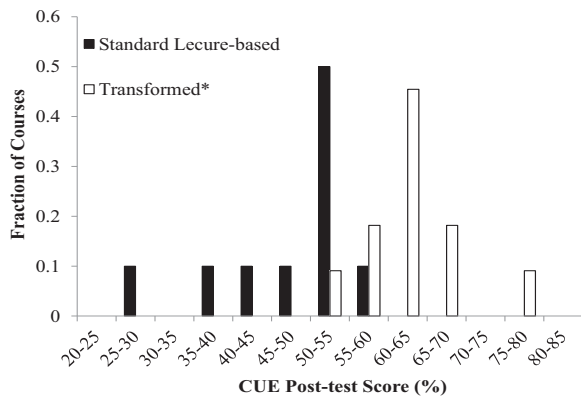}
\caption{Histogram of average course score on the CUE across 7 institutions demonstrating improved performance for courses using CU's transformed materials (11 courses, 9 CU; 329 students, 312 CU) relative to courses using only traditional lecture (10 courses, 5 CU; 303 students, 191 CU).  \\ * All transformed courses used some or all of CU's transformed electrostatics materials}\label{fig:CUEhake}
\end{figure}

For example, data on average scores on the CUE across 21 courses and 7 institutions demonstrate that transformed electrostatics courses score significantly higher on the CUE post-test (see Fig.\ \ref{fig:CUEhake}).  Using students as data points, transformed courses averaged 56.6 $\pm$ 1\% and traditional courses scored 45.7 $\pm$ 1\%.  Treating courses as data points, these averages shift to 58.0 $\pm$ 2\% and 42.3 $\pm$ 3\% respectively.  

\begin{figure}
    \includegraphics[height=2.1in, width=3.1in]{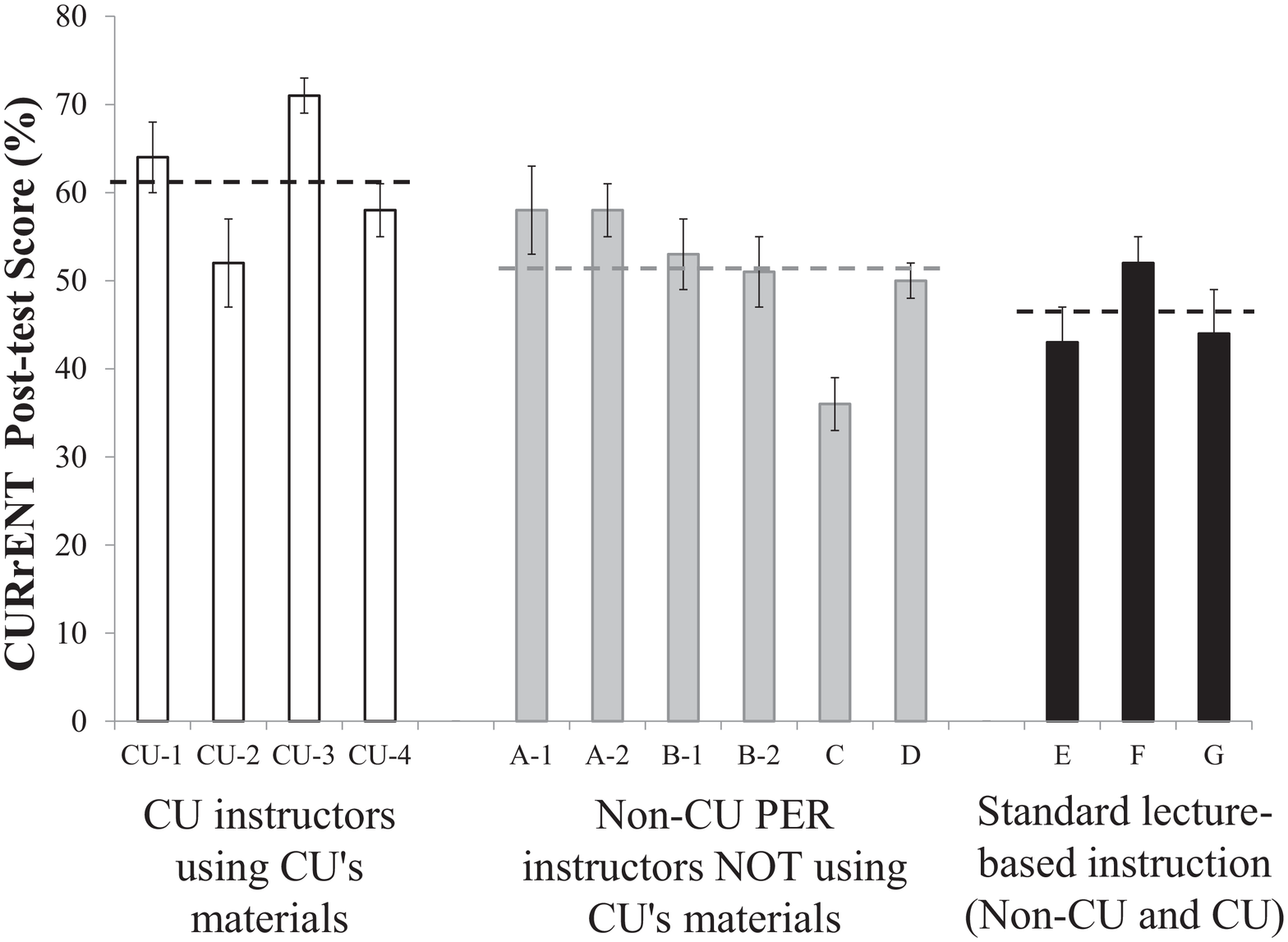}
\caption{Histogram of average course score on the CURrENT across 6 institutions demonstrating preliminary indications of improved performance for CU's transformed (N=124) and PER (N=172) courses relative to courses using only traditional lecture (N=50).  Overall averages by course are marked by the dashed lines.  }\label{fig:CURrENT}
\end{figure}

While we have considerably less data available from the CURrENT than the CUE, scores from 13 courses at 6 institutions also show preliminary indications that transformed curricular materials improve student learning as measured by the CURrENT (see Fig.\ \ref{fig:CURrENT}). Treating the courses as data points, CU transformed courses average 61 $\pm$ 4\% and courses taught by PER instructors but not using CU's transformed course materials average 51 $\pm$ 3\%.  This represents consistent improvement when compared with an average of 46 $\pm$ 3\% from courses taught using only traditional lecture.  However, the standard-lecture based sample in these data is small and more data collection will be necessary to more robustly establish the impact of our course transformations on student learning.  

While a discussion of the effectiveness of CU's transformed curricular materials is not the goal of this paper, it is worth noting a potential concern that we are `teaching to the test'.  Given that both our instruments and our transformed materials were designed to address our explicit learning goals, it is perhaps not surprising that students using our materials score higher on the assessments.   However, faculty at CU and elsewhere, including those using standard lecture-based instruction, have consistently indicated that they agree that our learning goals represent what they want students to know, and that the assessments target a subset of these learning goals. Our assessments reflect the concepts and goals that upper-division instructors value and provide information about how well their students understand those concepts and achieve those goals.


\subsection{\label{sec:administration}For administrative purposes}

Student performance on these instruments can also be used for administrative purposes.  For example, comparative data on student learning using different pedagogies have been a necessary (though not sufficient) condition in supporting faculty at CU and elsewhere in their efforts to incorporate interactive engagement techniques into their upper-division courses, and to sustaining the use of our transformed curricular materials at CU \cite{chasteen2011sustaining}.  It has also become common at CU for instructors to include their students' performance on these instruments as part of their tenure and promotion cases as evidence of reflective teaching practices.   Additionally, the CU Physics Department has presented these assessments to deans and potential donors as evidence of improvements to undergraduate STEM education at CU.  We have also been told informally that scores have been included in annual departmental reports at other institutions.

\subsection{\label{sec:difficulties}To investigate student difficulties}

In addition to using scores as comparative measures, our assessments have also been used to gain insight into the nature of common student difficulties.  The free-response format provides a particularly rich data source for identifying and characterizing topics that students find particularly challenging.  Examples of this from CU span a number of topics including: Gauss's Law \cite{pepper2010gauss}, Ampere's Law \cite{wallace2010ampere}, divergence of vector fields \cite{astolfi2014divergence}, Taylor series \cite{caballero2013ccmi, wilcox2013acer}, quantum energies/time development \cite{goldhaber2009qmat}, and electric potential \cite{pepper2012math}.  In some cases, investigation of difficulties identified by one of these assessments has inspired broader research efforts.  For example, early classroom testing of the CCMI revealed that our students often struggled not only with \emph{how} to use Taylor Series, but also knowing \emph{when} they were appropriate.  Subsequent investigation of this difficulty helped to inform the development of an analytical framework for student use of mathematical tools in physics that specifically attends to how students determine which tool is appropriate \cite{wilcox2013acer}.  

As an example of this insight into student difficulties, analysis of preliminary data from the QMAT showed that our students had significant difficulties with the relationship between the Hamiltonian and the time evolution of quantum states similar to those reported previously \cite{gire2008Qmath}.  Roughly half of these students agreed with the statement that applying the Hamiltonian to an arbitrary state gives information on how the state will evolve in time (QMAT Q4); however, only a quarter could also justify why the statement was true \cite{goldhaber2009qmat}.  Many students who disagreed with this statement focused on the lack of time dependence in the Hamiltonian itself.  Similarly, only a third of our students both disagreed with the statement that a system in an eigenstate of an arbitrary operator would stay in that state until disturbed (QMAT Q12), and could also justify why it was incorrect.  Data from the QMCA also suggest that the concepts of time evolution are particularly challenging, and that the difficult aspect may arise, in part, from an over-generalization of the unique aspects of energy measurements to physical observables whose corresponding operators do not commute with the Hamiltonian.  

In addition to providing insight into the nature of student difficulties, standardized conceptual assessments can also be used to determine and to compare the relative prevalence of these difficulties across institutions and pedagogies.  The QMCA provides a striking example of this.  Questions on both the QMCA and QMAT can be grouped into five main concept frames: measurement, the time independent Schr\"{o}dinger equation, wavefunctions and boundary conditions, time evolution, and probability/probability density.  The overall patterns of students' QMCA scores on these five topics are strikingly similar across 10 institutions.  The greatest variation appears in students' scores related to quantum measurement, whereas scores in the other four categories are practically the same \cite{sadaghiani2014qmca}.  This observation supports the existence of several wide-spread common student difficulties regardless of student population and type of institution.  

There are additional examples of our assessments being used to compare the prevalence of student difficulties between institutions.  Researchers at Oregon State University (OSU) have looked at responses to a subset of CUE questions from students at both OSU and CU to identify ways in which the two student populations differ in terms of both scores and prevalent difficulties \cite{zwolak2013CUErubric}.  In particular, they examined one question on the CUE that is most easily solved using superposition of either the electric field or potential.  They found that at both institutions, students often did not identify superposition as the correct solution method, or explicitly referred to the superposition of charges instead of fields.  However, they also found that students at OSU were less likely to use the term `superposition', and were more likely to use the superposition of electric potential than students at CU.  These differences likely reflect differences between the CU and OSU curriculum, as the OSU curriculum does not emphasize the term `superposition' and presents the electric potential before the electric field \cite{paradigms}.

\section{\label{sec:implementation}Implementation of CU's Upper-division Assessments}

\subsection{\label{sec:barriers} Barriers and Challenges}

Since their development, all of our upper-division assessments have been administered at multiple universities in the US; however, we have encountered a number of barriers and challenges to consistent use of these assessments on a large scale.  One barrier to large-scale implementation is faculty/instructor resistance to the assessments themselves.  As standardized assessment is not a normal part of upper-division physics instruction, some instructors are hesitant to give an assessment that could reflect poorly on their instruction or be used to set one faculty member up against others.  

Faculty can also be discouraged by the logistical requirements of these assessments.  Instructors must dedicate class time to give these assessments, typically a full 50 minute class period at the end of the semester. This can seem particularly onerous if the instructor is being asked to give the assessment by an outside source (e.g., a department chair or PER researcher).  Once given, the free-response format of our assessment also makes them challenging and time consuming to grade.  Busy faculty are often unable to dedicate the necessary time to grade these assessments.  

Instructors who administer these assessments can also experience some resistance from students.  For example, students often want to use these assessments as a study tool.  However, because these assessments are difficult and time consuming to develop, keeping them secure is particularly important.  For this reason, we actively discourage instructors from providing solutions for their students or allowing them to take the test with them.  Students can find these restrictions frustrating, particularly if no opportunity is given for them to review and ask questions about the assessment.  

Another challenge we have encountered both with the development and large-scale implementation of our upper-division assessments is the relative lack of consistency between the content coverage and pace of advanced physics courses compared to introductory courses.  The exact content of the upper-division physics curriculum can vary significantly from institution to institution and even from instructor to instructor.  It is also not unusual for instructors to feel more ownership of these advanced courses and thus there is a greater degree of customization of each course.  This makes it difficult to create a one-size-fits-all assessment that accurately reflects the content coverage and emphasis of the majority of courses.  While we believe our assessments are representative of broader courses in that they were designed to match canonical textbooks and consensus learning goals, some external institutions have argued that the instruments favor the particular content and teaching styles at CU \cite{zwolak2013CUErubric}. 

\subsection{\label{sec:solutions}Strategies and Solutions}

We have implemented a number of strategies to minimize the barriers and challenges documented above.  To minimize faculty resistance to the assessments themselves, we have solicited faculty involvement early in the development process to ensure they have the opportunity to help shape the instruments so that they value student outcomes on these measures.  To reduce some of the logistical barriers, we have consistently offered to help faculty with grading each of these assessments.  As a more sustainable strategy, our newer assessments (the CCMI and CURrENT) were both explicitly designed to have simple grading rubrics that are fast and straightforward to use.  This helps to minimize faculty concern about being able to grade these assessments.  Even more recently, we have begun developing multiple-choice and multiple-response versions of these assessments that allow for fast and objective grading. To date, the CUE and QMAT have been converted into two different easily-gradable formats; detailed discussion of these new versions can be found in Refs. \cite{wilcox2013cue, wilcox2014cmr, sadaghiani2013qmca, sadaghiani2014qmca}

We have also developed strategies to minimize student resistance.  Framing the tests as valuable, but low stakes measures of students' understanding that can be used to help them prepare for the final exam can be effective at promoting student buy-in.  When possible, we also provide individualized feedback for each student, which includes their overall score relative to the class average.  Additionally, offering a few extra office hours the final week of classes where students can come discuss and review their exams (without taking them home) can also help to encourage students to see these instruments as useful preparation for the final.  

Variable content coverage between courses is a more challenging barrier to address as it is in many ways a characteristic of upper-division physics instruction, rather than the assessments themselves.  However, this issue was particularly important for the CCMI, as the classical mechanics course at CU is a joint math methods course as well.  To address this, the CCMI includes two optional questions in addition to the 9 core questions.  These optional questions target several of the mathematical methods emphasized in the CU course but are not included in the score on the assessment because they are not representative of broader classical mechanics courses.  The issue of variable content coverage was also addressed early in the development of the CURrENT during a summer working group in which faculty from external institutions participated in discussions concerning the appropriate scope for the instrument.  This greater insight into what the E\&M 2 course looked like at other institutions directly motivated several restrictions in the content coverage of the CURrENT.

\section{\label{sec:conclusions}Summary \& Discussion}

Over the past decade, an increasing number of standardized and validated conceptual assessments have been developed that specifically target physics content beyond the introductory level.  Specific topic areas include classical mechanics, electricity and magnetism, quantum mechanics and thermodynamics.  In this paper, we identified and briefly compared many of these assessment instruments based on format, content coverage, and development.  We then provided a more detailed review of four instruments created at CU as an example of the development, validation, and uses of upper-division conceptual assessments in physics.  We also discussed some of the barriers to implementing these assessments in the classroom as well as some strategies and solutions to overcoming these barriers. 

Of the published assessment instruments discussed here, all were developed using a similar iterative design cycle involving initial development, expert reviews, student interviews, and preliminary classroom testing (Fig.\ \ref{fig:dCycle}).  At CU, the initial development phase was heavily influenced by consensus learning goals that emphasized more meta-level outcomes related to `thinking-like-a-physicist.'  These meta-level goals were the primary motivation for the unique free-response format of CU's assessments.  Initial development of other assessments (Table \ref{tab:instruments}) focused instead on achieving appropriate content coverage without explicit discussion of non-content related learning goals.  

Consistent with the literature published at the introductory level, the majority of the upper-division conceptual assessments described here (Table \ref{tab:instruments}) were validated using Classical Test Theory, and test statistics are available for all but two (the EMCI and QMAT).  In all cases, the majority of the statistics for a given instrument fell within accepted ranges, indicating that each offers a valid and reliable measure of student learning within the tested populations and contexts.  

There are a variety of examples of the uses of these assessment tools: as comparative measures of student learning across instructors, institutions, and time; and as sources of insight into student difficulties.  The latter use is particularly true for the four open-ended assessments from CU, as the free-response format allows for generation and identification of new student difficulties rather than primarily providing data on the prevalence of known student difficulties, as on a multiple-choice instrument.  However, we have also encountered a number of barriers to both small and large-scale implementation of conceptual assessments in the upper-division including: faculty resistance, student resistance, and logistical constraints.  In some cases, we have implemented strategies to reduce these barriers (e.g., creating simple grading rubrics and multiple-choice versions to simplify the grading process).  

While many of the barriers to conceptual assessment at the upper-division level are also, at least to some extent, barriers at the introductory level, one issue that is particularly acute at the advanced undergraduate level is the issue of variable course coverage.  Reduced consistency in content coverage and emphasis between instructors and institutions makes it very challenging to create assessment instruments that are appropriate for a broad range of courses.  This is reflected in, for example, the relatively large number of assessments available for advanced quantum mechanics, each with slightly different focus and scope.  Barring a national standardization of the upper-division physics curriculum, which we see as unlikely and potentially undesirable, one potential solution to this issue, requiring large-scale coordination of both the PER and broader physics communities, would be to create banks of questions that can be used by individual instructors to craft course-appropriate assessments.  This strategy is similar to what has been done for large-scale testing in K-12 (i.e., SAT or ACT testing) and would require the use of Item Response Theory to validate all potential items.  

Ongoing work with CU's upper-division assessments includes completing final classroom tests of the CCMI and CURrENT, as well as the CMR CUE and QMCA.  Particular emphasis is being placed on expanding classroom testing beyond the developing institution in order to more robustly establish the validity of these assessments for a broader spectrum of physics students.  Future work may include leveraging these assessments as longitudinal measures of student learning, creating new assessments for additional topical areas (e.g., thermodyamics), and/or converting the CCMI and CURrENT to multiple-choice or multiple-response formats to further facilitate large-scale use.  

The translation of CU's free-response assessments into multiple-choice/multiple-response versions was motivated entirely by a desire to increase the scalability and usability, and is not an indication that we see the free-response versions as obsolete.  The logistical advantages of the multiple-choice formats come with significant trade-offs (e.g., reduced insight into details of student thinking and exclusion of unanticipated responses).  Ultimately, which version of the assessment should be used in any given context is dependent on both the kind of information an instructor or researcher wants to capture (e.g., comparative scores vs. deeper insight into student reasoning) as well as the logistical constraints of the specific course/program (e.g., class size).  Thus, there is value in having both formats available for use in different contexts.  

Even with an explicit emphasis on CU's meta-level learning goals, upper-division conceptual assessments are still heavily content focused.  Yet, there are many skills and characteristics related to a student's development as a physicist that extend beyond content knowledge that are rarely, if ever, assessed directly.  For example, capacity for independent learning, ability to read and write scientific publications, and ability to work collaboratively are just a few characteristics of successful physicists that we ultimately want our physics majors to internalize.  We argue that operationalizing and assessing these implicit goals represents an important outstanding issue for the PER community to consider.  Can we begin to craft assessments that more accurately reflect the full range of learning outcomes we value for our physics majors?

\begin{acknowledgments}
Particular thanks to the PER@C group and the faculty and students who contributed to the development of these assessments.
This work was funded by the Science Education Initiative, an NSF-CCLI Grant DUE-1023028 and an NSF Graduate Research Fellowship under Grant No. DGE 1144083.
\end{acknowledgments}

\bibliography{master-refs}

\begin{thebibliography}{50}
\expandafter\ifx\csname natexlab\endcsname\relax\def\natexlab#1{#1}\fi
\expandafter\ifx\csname bibnamefont\endcsname\relax
  \def\bibnamefont#1{#1}\fi
\expandafter\ifx\csname bibfnamefont\endcsname\relax
  \def\bibfnamefont#1{#1}\fi
\expandafter\ifx\csname citenamefont\endcsname\relax
  \def\citenamefont#1{#1}\fi
\expandafter\ifx\csname url\endcsname\relax
  \def\url#1{\texttt{#1}}\fi
\expandafter\ifx\csname urlprefix\endcsname\relax\def\urlprefix{URL }\fi
\providecommand{\bibinfo}[2]{#2}
\providecommand{\eprint}[2][]{\url{#2}}

\bibitem[{\citenamefont{Hestenes et~al.}(1992)\citenamefont{Hestenes, Wells,
  and Swackhamer}}]{hestenes1992fci}
\bibinfo{author}{\bibfnamefont{D.}~\bibnamefont{Hestenes}},
  \bibinfo{author}{\bibfnamefont{M.}~\bibnamefont{Wells}}, \bibnamefont{and}
  \bibinfo{author}{\bibfnamefont{G.}~\bibnamefont{Swackhamer}},
  \emph{\bibinfo{title}{Force concept inventory}}, \bibinfo{journal}{The
  Physics Teacher} \textbf{\bibinfo{volume}{30}}, \bibinfo{pages}{141}
  (\bibinfo{year}{1992}),
  \urlprefix\url{http://link.aip.org/link/?PTE/30/141/1}.

\bibitem[{\citenamefont{Thornton and Sokoloff}(1998)}]{thornton1998fmce}
\bibinfo{author}{\bibfnamefont{R.~K.} \bibnamefont{Thornton}} \bibnamefont{and}
  \bibinfo{author}{\bibfnamefont{D.~R.} \bibnamefont{Sokoloff}},
  \emph{\bibinfo{title}{Assessing student learning of newton’s laws: The
  force and motion conceptual evaluation and the evaluation of active learning
  laboratory and lecture curricula}}, \bibinfo{journal}{Am. J. Phys.}
  \textbf{\bibinfo{volume}{66}} (\bibinfo{year}{1998}).

\bibitem[{\citenamefont{Ding et~al.}(2006)\citenamefont{Ding, Chabay, Sherwood,
  and Beichner}}]{ding2006bema}
\bibinfo{author}{\bibfnamefont{L.}~\bibnamefont{Ding}},
  \bibinfo{author}{\bibfnamefont{R.}~\bibnamefont{Chabay}},
  \bibinfo{author}{\bibfnamefont{B.}~\bibnamefont{Sherwood}}, \bibnamefont{and}
  \bibinfo{author}{\bibfnamefont{R.}~\bibnamefont{Beichner}},
  \emph{\bibinfo{title}{Evaluating an electricity and magnetism assessment
  tool: Brief electricity and magnetism assessment}}, \bibinfo{journal}{Phys.
  Rev. ST Phys. Educ. Res.} \textbf{\bibinfo{volume}{2}},
  \bibinfo{pages}{010105} (\bibinfo{year}{2006}),
  \urlprefix\url{http://link.aps.org/doi/10.1103/PhysRevSTPER.2.010105}.

\bibitem[{\citenamefont{{http://www.ncsu.edu/per/TestInfo.html}}(2014)}]{beichnerAssessment}
\bibinfo{author}{\bibnamefont{{http://www.ncsu.edu/per/TestInfo.html}}}
  (\bibinfo{year}{2014}).

\bibitem[{\citenamefont{Hake}(1998)}]{hake1998ie}
\bibinfo{author}{\bibfnamefont{R.~R.} \bibnamefont{Hake}},
  \emph{\bibinfo{title}{Interactive-engagement versus traditional methods: A
  six-thousand-student survey of mechanics test data for introductory physics
  courses}}, \bibinfo{journal}{Am. J. Phys.} \textbf{\bibinfo{volume}{66}},
  \bibinfo{pages}{64} (\bibinfo{year}{1998}),
  \urlprefix\url{http://link.aip.org/link/?AJP/66/64/1}.

\bibitem[{\citenamefont{Chasteen
  et~al.}(2012{\natexlab{a}})\citenamefont{Chasteen, Pollock, Pepper, and
  Perkins}}]{chasteen2012transforming}
\bibinfo{author}{\bibfnamefont{S.~V.} \bibnamefont{Chasteen}},
  \bibinfo{author}{\bibfnamefont{S.~J.} \bibnamefont{Pollock}},
  \bibinfo{author}{\bibfnamefont{R.~E.} \bibnamefont{Pepper}},
  \bibnamefont{and} \bibinfo{author}{\bibfnamefont{K.~K.}
  \bibnamefont{Perkins}}, \emph{\bibinfo{title}{Transforming the junior level:
  Outcomes from instruction and research in e\&m}}, \bibinfo{journal}{Phys.
  Rev. ST Phys. Educ. Res.} \textbf{\bibinfo{volume}{8}},
  \bibinfo{pages}{020107} (\bibinfo{year}{2012}{\natexlab{a}}).

\bibitem[{\citenamefont{Caballero and Pollock}(2013)}]{caballero2013ccmi}
\bibinfo{author}{\bibfnamefont{M.}~\bibnamefont{Caballero}} \bibnamefont{and}
  \bibinfo{author}{\bibfnamefont{S.}~\bibnamefont{Pollock}}, in
  \emph{\bibinfo{booktitle}{Physics Education Research Conference 2013}}
  (\bibinfo{address}{Portland, OR}, \bibinfo{year}{2013}), PER Conference, pp.
  \bibinfo{pages}{81--84}.

\bibitem[{\citenamefont{Notaros}(2002)}]{notaros2002emci}
\bibinfo{author}{\bibfnamefont{B.~M.} \bibnamefont{Notaros}}, in
  \emph{\bibinfo{booktitle}{Antennas and Propagation Society International
  Symposium, 2002. IEEE}} (\bibinfo{organization}{IEEE}, \bibinfo{year}{2002}),
  vol.~\bibinfo{volume}{1}, pp. \bibinfo{pages}{684--687}.

\bibitem[{\citenamefont{Chasteen
  et~al.}(2012{\natexlab{b}})\citenamefont{Chasteen, Pepper, Caballero,
  Pollock, and Perkins}}]{chasteen2012cue}
\bibinfo{author}{\bibfnamefont{S.~V.} \bibnamefont{Chasteen}},
  \bibinfo{author}{\bibfnamefont{R.~E.} \bibnamefont{Pepper}},
  \bibinfo{author}{\bibfnamefont{M.~D.} \bibnamefont{Caballero}},
  \bibinfo{author}{\bibfnamefont{S.~J.} \bibnamefont{Pollock}},
  \bibnamefont{and} \bibinfo{author}{\bibfnamefont{K.~K.}
  \bibnamefont{Perkins}}, \emph{\bibinfo{title}{Colorado upper-division
  electrostatics diagnostic: A conceptual assessment for the junior level}},
  \bibinfo{journal}{Phys. Rev. ST Phys. Educ. Res.}
  \textbf{\bibinfo{volume}{8}}, \bibinfo{pages}{020108}
  (\bibinfo{year}{2012}{\natexlab{b}}).

\bibitem[{\citenamefont{Baily et~al.}(2012)\citenamefont{Baily, Dubson, and
  Pollock}}]{baily2012materials}
\bibinfo{author}{\bibfnamefont{C.}~\bibnamefont{Baily}},
  \bibinfo{author}{\bibfnamefont{M.}~\bibnamefont{Dubson}}, \bibnamefont{and}
  \bibinfo{author}{\bibfnamefont{S.}~\bibnamefont{Pollock}}, in
  \emph{\bibinfo{booktitle}{Physics Education Research Conference 2012}}
  (\bibinfo{address}{Philadelphia, PA}, \bibinfo{year}{2012}), vol.
  \bibinfo{volume}{1513} of \emph{\bibinfo{series}{PER Conference}}, pp.
  \bibinfo{pages}{54--57}.

\bibitem[{\citenamefont{Zhu and Singh}(2012)}]{zhu2012qms}
\bibinfo{author}{\bibfnamefont{G.}~\bibnamefont{Zhu}} \bibnamefont{and}
  \bibinfo{author}{\bibfnamefont{C.}~\bibnamefont{Singh}},
  \emph{\bibinfo{title}{Surveying students' understanding of quantum mechanics
  in one spatial dimension}}, \bibinfo{journal}{Am. J. Phys.}
  \textbf{\bibinfo{volume}{80}}, \bibinfo{pages}{252} (\bibinfo{year}{2012}).

\bibitem[{\citenamefont{Cataloglu and Robinett}(2002)}]{cataloglu2002qmvi}
\bibinfo{author}{\bibfnamefont{E.}~\bibnamefont{Cataloglu}} \bibnamefont{and}
  \bibinfo{author}{\bibfnamefont{R.}~\bibnamefont{Robinett}},
  \emph{\bibinfo{title}{Testing the development of student conceptual and
  visualization understanding in quantum mechanics through the undergraduate
  career}}, \bibinfo{journal}{American Journal of Physics}
  \textbf{\bibinfo{volume}{70}}, \bibinfo{pages}{238} (\bibinfo{year}{2002}).

\bibitem[{\citenamefont{Goldhaber et~al.}(2009)\citenamefont{Goldhaber,
  Pollock, Dubson, Beale, and Perkins}}]{goldhaber2009qmat}
\bibinfo{author}{\bibfnamefont{S.}~\bibnamefont{Goldhaber}},
  \bibinfo{author}{\bibfnamefont{S.}~\bibnamefont{Pollock}},
  \bibinfo{author}{\bibfnamefont{M.}~\bibnamefont{Dubson}},
  \bibinfo{author}{\bibfnamefont{P.}~\bibnamefont{Beale}}, \bibnamefont{and}
  \bibinfo{author}{\bibfnamefont{K.}~\bibnamefont{Perkins}}, in
  \emph{\bibinfo{booktitle}{Physics Education Research Conference 2009}}
  (\bibinfo{address}{Ann Arbor, Michigan}, \bibinfo{year}{2009}), vol.
  \bibinfo{volume}{1179} of \emph{\bibinfo{series}{PER Conference}}, pp.
  \bibinfo{pages}{145--148}.

\bibitem[{\citenamefont{Midkiff et~al.}(2001)\citenamefont{Midkiff, Litzinger,
  and Evans}}]{midkiff2001tci}
\bibinfo{author}{\bibfnamefont{K.~C.} \bibnamefont{Midkiff}},
  \bibinfo{author}{\bibfnamefont{T.~A.} \bibnamefont{Litzinger}},
  \bibnamefont{and} \bibinfo{author}{\bibfnamefont{D.}~\bibnamefont{Evans}}, in
  \emph{\bibinfo{booktitle}{Frontiers in Education Conference, 2001. 31st
  Annual}} (\bibinfo{organization}{IEEE}, \bibinfo{year}{2001}),
  vol.~\bibinfo{volume}{2}, pp. \bibinfo{pages}{F2A--F23}.

\bibitem[{\citenamefont{Streveler et~al.}(2011)\citenamefont{Streveler, Miller,
  Santiago-Rom{\'a}n, Nelson, Geist, and Olds}}]{streveler2011ttci}
\bibinfo{author}{\bibfnamefont{R.~A.} \bibnamefont{Streveler}},
  \bibinfo{author}{\bibfnamefont{R.~L.} \bibnamefont{Miller}},
  \bibinfo{author}{\bibfnamefont{A.~I.} \bibnamefont{Santiago-Rom{\'a}n}},
  \bibinfo{author}{\bibfnamefont{M.~A.} \bibnamefont{Nelson}},
  \bibinfo{author}{\bibfnamefont{M.~R.} \bibnamefont{Geist}}, \bibnamefont{and}
  \bibinfo{author}{\bibfnamefont{B.~M.} \bibnamefont{Olds}},
  \emph{\bibinfo{title}{Rigorous methodology for concept inventory development:
  Using the'assessment triangle'to develop and test the thermal and transport
  science concept inventory (ttci)}}, \bibinfo{journal}{International Journal
  of Engineering Education} \textbf{\bibinfo{volume}{27}}, \bibinfo{pages}{968}
  (\bibinfo{year}{2011}).

\bibitem[{\citenamefont{Evans et~al.}(2003)\citenamefont{Evans, Gray, Krause,
  Martin, Midkiff, Notaros, Pavelich, Rancour, Reed-Rhoads, Steif
  et~al.}}]{evans2003CIs}
\bibinfo{author}{\bibfnamefont{D.}~\bibnamefont{Evans}},
  \bibinfo{author}{\bibfnamefont{G.~L.} \bibnamefont{Gray}},
  \bibinfo{author}{\bibfnamefont{S.}~\bibnamefont{Krause}},
  \bibinfo{author}{\bibfnamefont{J.}~\bibnamefont{Martin}},
  \bibinfo{author}{\bibfnamefont{C.}~\bibnamefont{Midkiff}},
  \bibinfo{author}{\bibfnamefont{B.~M.} \bibnamefont{Notaros}},
  \bibinfo{author}{\bibfnamefont{M.}~\bibnamefont{Pavelich}},
  \bibinfo{author}{\bibfnamefont{D.}~\bibnamefont{Rancour}},
  \bibinfo{author}{\bibfnamefont{T.}~\bibnamefont{Reed-Rhoads}},
  \bibinfo{author}{\bibfnamefont{P.}~\bibnamefont{Steif}},
  \bibnamefont{et~al.}, in \emph{\bibinfo{booktitle}{Frontiers in Education,
  2003. FIE 2003 33rd Annual}} (\bibinfo{organization}{IEEE},
  \bibinfo{year}{2003}), vol.~\bibinfo{volume}{1}, pp. \bibinfo{pages}{T4G--1}.

\bibitem[{\citenamefont{Thoads and Roedel}(1999)}]{thoads1999wave}
\bibinfo{author}{\bibfnamefont{T.}~\bibnamefont{Thoads}} \bibnamefont{and}
  \bibinfo{author}{\bibfnamefont{R.~J.} \bibnamefont{Roedel}}, in
  \emph{\bibinfo{booktitle}{Frontiers in Education Conference, 1999. FIE'99.
  29th Annual}} (\bibinfo{organization}{IEEE}, \bibinfo{year}{1999}),
  vol.~\bibinfo{volume}{3}, pp. \bibinfo{pages}{13C1--14}.

\bibitem[{\citenamefont{Singh}(2006)}]{singh2006scge}
\bibinfo{author}{\bibfnamefont{C.}~\bibnamefont{Singh}},
  \emph{\bibinfo{title}{Student understanding of symmetry and gauss’s law of
  electricity}}, \bibinfo{journal}{Am. J. Phys.} \textbf{\bibinfo{volume}{74}}
  (\bibinfo{year}{2006}).

\bibitem[{\citenamefont{McKagan et~al.}(2010)\citenamefont{McKagan, Perkins,
  and Wieman}}]{mckagan2010qmcs}
\bibinfo{author}{\bibfnamefont{S.~B.} \bibnamefont{McKagan}},
  \bibinfo{author}{\bibfnamefont{K.~K.} \bibnamefont{Perkins}},
  \bibnamefont{and} \bibinfo{author}{\bibfnamefont{C.~E.}
  \bibnamefont{Wieman}}, \emph{\bibinfo{title}{Design and validation of the
  quantum mechanics conceptual survey}}, \bibinfo{journal}{Phys. Rev. ST Phys.
  Educ. Res.} \textbf{\bibinfo{volume}{6}}, \bibinfo{pages}{020121}
  (\bibinfo{year}{2010}),
  \urlprefix\url{http://link.aps.org/doi/10.1103/PhysRevSTPER.6.020121}.

\bibitem[{\citenamefont{Taylor}(2005)}]{taylor2005classmech}
\bibinfo{author}{\bibfnamefont{J.~R.} \bibnamefont{Taylor}},
  \emph{\bibinfo{title}{Classical mechanics}} (\bibinfo{publisher}{University
  Science Books}, \bibinfo{year}{2005}), ISBN \bibinfo{isbn}{9781891389221},
  \urlprefix\url{http://books.google.com/books?id=P1kCtNr-pJsC}.

\bibitem[{\citenamefont{Griffiths}(1999)}]{griffiths1999em}
\bibinfo{author}{\bibfnamefont{D.~J.} \bibnamefont{Griffiths}},
  \emph{\bibinfo{title}{Introduction to electrodynamics}}
  (\bibinfo{publisher}{Prentice Hall}, \bibinfo{year}{1999}), ISBN
  \bibinfo{isbn}{9780138053260},
  \urlprefix\url{http://books.google.com/books?id=M8XvAAAAMAAJ}.

\bibitem[{\citenamefont{{Ryan} et~al.}(2014)\citenamefont{{Ryan}, {Astolfi},
  {Baily}, and {Pollock}}}]{ryan2014current}
\bibinfo{author}{\bibfnamefont{Q.}~\bibnamefont{{Ryan}}},
  \bibinfo{author}{\bibfnamefont{C.}~\bibnamefont{{Astolfi}}},
  \bibinfo{author}{\bibfnamefont{C.}~\bibnamefont{{Baily}}}, \bibnamefont{and}
  \bibinfo{author}{\bibfnamefont{S.}~\bibnamefont{{Pollock}}}, in
  \emph{\bibinfo{booktitle}{Physics Education Research Conference 2014}}
  (\bibinfo{address}{Minneapolis, MN}, \bibinfo{year}{2014}), vol.
  \bibinfo{volume}{In press} of \emph{\bibinfo{series}{PER Conference}}.

\bibitem[{\citenamefont{Griffiths and Harris}(1995)}]{griffiths1995qm}
\bibinfo{author}{\bibfnamefont{D.~J.} \bibnamefont{Griffiths}}
  \bibnamefont{and} \bibinfo{author}{\bibfnamefont{E.~G.}
  \bibnamefont{Harris}}, \emph{\bibinfo{title}{Introduction to quantum
  mechanics}}, vol.~\bibinfo{volume}{2} (\bibinfo{publisher}{Prentice Hall New
  Jersey}, \bibinfo{year}{1995}).

\bibitem[{\citenamefont{Marion and Thornton}(2003)}]{marion2003classical}
\bibinfo{author}{\bibfnamefont{J.~B.} \bibnamefont{Marion}} \bibnamefont{and}
  \bibinfo{author}{\bibfnamefont{S.~T.} \bibnamefont{Thornton}},
  \emph{\bibinfo{title}{Classical dynamics of particles and systems}}
  (\bibinfo{publisher}{Brooks/Cole Cengage Learning}, \bibinfo{year}{2003}).

\bibitem[{\citenamefont{Pepper et~al.}(2011)\citenamefont{Pepper, Chasteen,
  Pollock, and Perkins}}]{pepper2011faculty}
\bibinfo{author}{\bibfnamefont{R.}~\bibnamefont{Pepper}},
  \bibinfo{author}{\bibfnamefont{S.}~\bibnamefont{Chasteen}},
  \bibinfo{author}{\bibfnamefont{S.}~\bibnamefont{Pollock}}, \bibnamefont{and}
  \bibinfo{author}{\bibfnamefont{K.}~\bibnamefont{Perkins}}, in
  \emph{\bibinfo{booktitle}{Physics Education Research Conference 2011}}
  (\bibinfo{address}{Omaha, Nebraska}, \bibinfo{year}{2011}), vol.
  \bibinfo{volume}{1413} of \emph{\bibinfo{series}{PER Conference}}, pp.
  \bibinfo{pages}{291--294}.

\bibitem[{\citenamefont{Chasteen et~al.}(2014)\citenamefont{Chasteen, Wilcox,
  Caballero, Perkins, Pollock, and Wieman}}]{chasteen2014sei}
\bibinfo{author}{\bibfnamefont{S.~V.} \bibnamefont{Chasteen}},
  \bibinfo{author}{\bibfnamefont{B.~R.} \bibnamefont{Wilcox}},
  \bibinfo{author}{\bibfnamefont{M.~D.} \bibnamefont{Caballero}},
  \bibinfo{author}{\bibfnamefont{K.~K.} \bibnamefont{Perkins}},
  \bibinfo{author}{\bibfnamefont{S.~J.} \bibnamefont{Pollock}},
  \bibnamefont{and} \bibinfo{author}{\bibfnamefont{C.}~\bibnamefont{Wieman}},
  \emph{\bibinfo{title}{Educational transformation in upper-division physics:
  The science education initiative model, outcomes, and lessons learned}},
  \bibinfo{journal}{Phys. Rev. ST Phys. Educ. Res.} \textbf{\bibinfo{volume}{In
  review}} (\bibinfo{year}{2014}).

\bibitem[{\citenamefont{{http://per.colorado.edu/sei}}(2014)}]{sei}
\bibinfo{author}{\bibnamefont{{http://per.colorado.edu/sei}}}
  (\bibinfo{year}{2014}).

\bibitem[{\citenamefont{{Doughty} and
  {Caballero}}(2014)}]{doughty2014ccmiDifficulties}
\bibinfo{author}{\bibfnamefont{L.}~\bibnamefont{{Doughty}}} \bibnamefont{and}
  \bibinfo{author}{\bibfnamefont{M.~D.} \bibnamefont{{Caballero}}}, in
  \emph{\bibinfo{booktitle}{Physics Education Research Conference 2014}}
  (\bibinfo{address}{Minneapolis, MN}, \bibinfo{year}{2014}), vol.
  \bibinfo{volume}{In press} of \emph{\bibinfo{series}{PER Conference}}.

\bibitem[{\citenamefont{{Wilcox} and {Pollock}}(2014)}]{wilcox2014cmr}
\bibinfo{author}{\bibfnamefont{B.~R.} \bibnamefont{{Wilcox}}} \bibnamefont{and}
  \bibinfo{author}{\bibfnamefont{S.~J.} \bibnamefont{{Pollock}}},
  \emph{\bibinfo{title}{{Coupled Multiple-response vs. Free-response Conceptual
  assessment: An Example from upper-division Physics}}},
  \bibinfo{journal}{Phys. Rev. ST Phys. Educ. Res.} \textbf{\bibinfo{volume}{In
  press}} (\bibinfo{year}{2014}).

\bibitem[{\citenamefont{Sadaghiani et~al.}(2013)\citenamefont{Sadaghiani,
  Miller, Pollock, and Rehn}}]{sadaghiani2013qmca}
\bibinfo{author}{\bibfnamefont{H.}~\bibnamefont{Sadaghiani}},
  \bibinfo{author}{\bibfnamefont{J.}~\bibnamefont{Miller}},
  \bibinfo{author}{\bibfnamefont{S.}~\bibnamefont{Pollock}}, \bibnamefont{and}
  \bibinfo{author}{\bibfnamefont{D.}~\bibnamefont{Rehn}}, in
  \emph{\bibinfo{booktitle}{Physics Education Research Conference 2013}}
  (\bibinfo{address}{Portland, OR}, \bibinfo{year}{2013}), PER Conference, pp.
  \bibinfo{pages}{319--322}.

\bibitem[{\citenamefont{Sadaghiani and Pollock}(2014)}]{sadaghiani2014qmca}
\bibinfo{author}{\bibfnamefont{H.}~\bibnamefont{Sadaghiani}} \bibnamefont{and}
  \bibinfo{author}{\bibfnamefont{S.}~\bibnamefont{Pollock}},
  \emph{\bibinfo{title}{Quantum mechanics conceptual assessment: Development
  and validation study}}, \bibinfo{journal}{Phys. Rev. ST Phys. Educ. Res.}
  \textbf{\bibinfo{volume}{Under review}} (\bibinfo{year}{2014}).

\bibitem[{\citenamefont{Engelhardt}(2009)}]{engelhardt2009ctt}
\bibinfo{author}{\bibfnamefont{P.}~\bibnamefont{Engelhardt}}, in
  \emph{\bibinfo{booktitle}{Getting Started in PER}} (\bibinfo{year}{2009}),
  vol.~\bibinfo{volume}{2}.

\bibitem[{\citenamefont{Ding and Beichner}(2009)}]{ding2009mcanalysis}
\bibinfo{author}{\bibfnamefont{L.}~\bibnamefont{Ding}} \bibnamefont{and}
  \bibinfo{author}{\bibfnamefont{R.}~\bibnamefont{Beichner}},
  \emph{\bibinfo{title}{Approaches to data analysis of multiple-choice
  questions}}, \bibinfo{journal}{Phys. Rev. ST Phys. Educ. Res.}
  \textbf{\bibinfo{volume}{5}}, \bibinfo{pages}{020103} (\bibinfo{year}{2009}).

\bibitem[{\citenamefont{Ding}(2014)}]{ding2014bemaVrasch}
\bibinfo{author}{\bibfnamefont{L.}~\bibnamefont{Ding}},
  \emph{\bibinfo{title}{Seeking missing pieces in science concept assessments:
  Reevaluating the brief electricity and magnetism assessment through rasch
  analysis}}, \bibinfo{journal}{Phys. Rev. ST Phys. Educ. Res}
  \textbf{\bibinfo{volume}{10}}, \bibinfo{pages}{010105}
  (\bibinfo{year}{2014}),
  \urlprefix\url{http://link.aps.org/doi/10.1103/PhysRevSTPER.10.010105}.

\bibitem[{\citenamefont{Aslanides and Savage}(2013)}]{aslanides2013rciVirt}
\bibinfo{author}{\bibfnamefont{J.~S.} \bibnamefont{Aslanides}}
  \bibnamefont{and} \bibinfo{author}{\bibfnamefont{C.~M.}
  \bibnamefont{Savage}}, \emph{\bibinfo{title}{Relativity concept inventory:
  Development, analysis, and results}}, \bibinfo{journal}{Phys. Rev. ST Phys.
  Educ. Res.} \textbf{\bibinfo{volume}{9}}, \bibinfo{pages}{010118}
  (\bibinfo{year}{2013}),
  \urlprefix\url{http://link.aps.org/doi/10.1103/PhysRevSTPER.9.010118}.

\bibitem[{\citenamefont{Marshall et~al.}(2009)\citenamefont{Marshall, Hagedorn,
  and O'Connor}}]{marshall2009taksVirt}
\bibinfo{author}{\bibfnamefont{J.~A.} \bibnamefont{Marshall}},
  \bibinfo{author}{\bibfnamefont{E.~A.} \bibnamefont{Hagedorn}},
  \bibnamefont{and} \bibinfo{author}{\bibfnamefont{J.}~\bibnamefont{O'Connor}},
  \emph{\bibinfo{title}{Anatomy of a physics test: Validation of the physics
  items on the texas assessment of knowledge and skills}},
  \bibinfo{journal}{Phys. Rev. ST Phys. Educ. Res.}
  \textbf{\bibinfo{volume}{5}}, \bibinfo{pages}{010104} (\bibinfo{year}{2009}),
  \urlprefix\url{http://link.aps.org/doi/10.1103/PhysRevSTPER.5.010104}.

\bibitem[{\citenamefont{Planinic et~al.}(2010)\citenamefont{Planinic, Ivanjek,
  and Susac}}]{planinic2010fciVrasch}
\bibinfo{author}{\bibfnamefont{M.}~\bibnamefont{Planinic}},
  \bibinfo{author}{\bibfnamefont{L.}~\bibnamefont{Ivanjek}}, \bibnamefont{and}
  \bibinfo{author}{\bibfnamefont{A.}~\bibnamefont{Susac}},
  \emph{\bibinfo{title}{Rasch model based analysis of the force concept
  inventory}}, \bibinfo{journal}{Phys. Rev. ST Phys. Educ. Res.}
  \textbf{\bibinfo{volume}{6}}, \bibinfo{pages}{010103} (\bibinfo{year}{2010}),
  \urlprefix\url{http://link.aps.org/doi/10.1103/PhysRevSTPER.6.010103}.

\bibitem[{\citenamefont{Wallace and Bailey}(2010)}]{wallace2010irt}
\bibinfo{author}{\bibfnamefont{C.~S.} \bibnamefont{Wallace}} \bibnamefont{and}
  \bibinfo{author}{\bibfnamefont{J.~M.} \bibnamefont{Bailey}},
  \emph{\bibinfo{title}{Do concept inventories actually measure anything?}},
  \bibinfo{journal}{Astronomy Education Review} \textbf{\bibinfo{volume}{9}},
  \bibinfo{pages}{010116} (\bibinfo{year}{2010}).

\bibitem[{\citenamefont{De~Ayala}(2009)}]{deayala2009polyIRT}
\bibinfo{author}{\bibfnamefont{R.~J.} \bibnamefont{De~Ayala}},
  \emph{\bibinfo{title}{Theory and practice of item response theory}}
  (\bibinfo{publisher}{Guilford Publications}, \bibinfo{year}{2009}).

\bibitem[{\citenamefont{Cortina}(1993)}]{cortina1993ca}
\bibinfo{author}{\bibfnamefont{J.~M.} \bibnamefont{Cortina}},
  \emph{\bibinfo{title}{What is coefficient alpha? an examination of theory and
  applications.}}, \bibinfo{journal}{Journal of applied psychology}
  \textbf{\bibinfo{volume}{78}}, \bibinfo{pages}{98} (\bibinfo{year}{1993}).

\bibitem[{\citenamefont{Chasteen et~al.}(2011)\citenamefont{Chasteen, Pepper,
  Pollock, and Perkins}}]{chasteen2011sustaining}
\bibinfo{author}{\bibfnamefont{S.}~\bibnamefont{Chasteen}},
  \bibinfo{author}{\bibfnamefont{R.}~\bibnamefont{Pepper}},
  \bibinfo{author}{\bibfnamefont{S.}~\bibnamefont{Pollock}}, \bibnamefont{and}
  \bibinfo{author}{\bibfnamefont{K.}~\bibnamefont{Perkins}}, in
  \emph{\bibinfo{booktitle}{Physics Education Research Conference 2011}}
  (\bibinfo{address}{Omaha, Nebraska}, \bibinfo{year}{2011}), vol.
  \bibinfo{volume}{1413} of \emph{\bibinfo{series}{PER Conference}}, pp.
  \bibinfo{pages}{139--142}.

\bibitem[{\citenamefont{Pepper et~al.}(2010)\citenamefont{Pepper, Chasteen,
  Pollock, and Perkins}}]{pepper2010gauss}
\bibinfo{author}{\bibfnamefont{R.}~\bibnamefont{Pepper}},
  \bibinfo{author}{\bibfnamefont{S.}~\bibnamefont{Chasteen}},
  \bibinfo{author}{\bibfnamefont{S.}~\bibnamefont{Pollock}}, \bibnamefont{and}
  \bibinfo{author}{\bibfnamefont{K.}~\bibnamefont{Perkins}}, in
  \emph{\bibinfo{booktitle}{Physics Education Research Conference 2010}}
  (\bibinfo{address}{Portland, Oregon}, \bibinfo{year}{2010}), vol.
  \bibinfo{volume}{1289} of \emph{\bibinfo{series}{PER Conference}}, pp.
  \bibinfo{pages}{245--248}.

\bibitem[{\citenamefont{Wallace and Chasteen}(2010)}]{wallace2010ampere}
\bibinfo{author}{\bibfnamefont{C.~S.} \bibnamefont{Wallace}} \bibnamefont{and}
  \bibinfo{author}{\bibfnamefont{S.~V.} \bibnamefont{Chasteen}},
  \emph{\bibinfo{title}{Upper-division students' difficulties with amp\`ere's
  law}}, \bibinfo{journal}{Phys. Rev. ST Phys. Educ. Res.}
  \textbf{\bibinfo{volume}{6}}, \bibinfo{pages}{020115} (\bibinfo{year}{2010}),
  \urlprefix\url{http://link.aps.org/doi/10.1103/PhysRevSTPER.6.020115}.

\bibitem[{\citenamefont{{Baily} and {Astolfi}}(2014)}]{astolfi2014divergence}
\bibinfo{author}{\bibfnamefont{C.}~\bibnamefont{{Baily}}} \bibnamefont{and}
  \bibinfo{author}{\bibfnamefont{C.}~\bibnamefont{{Astolfi}}}, in
  \emph{\bibinfo{booktitle}{Physics Education Research Conference 2014}}
  (\bibinfo{address}{Minneapolis, MN}, \bibinfo{year}{2014}), vol.
  \bibinfo{volume}{In press} of \emph{\bibinfo{series}{PER Conference}}.

\bibitem[{\citenamefont{Wilcox et~al.}(2013)\citenamefont{Wilcox, Caballero,
  Rehn, and Pollock}}]{wilcox2013acer}
\bibinfo{author}{\bibfnamefont{B.~R.} \bibnamefont{Wilcox}},
  \bibinfo{author}{\bibfnamefont{M.~D.} \bibnamefont{Caballero}},
  \bibinfo{author}{\bibfnamefont{D.~A.} \bibnamefont{Rehn}}, \bibnamefont{and}
  \bibinfo{author}{\bibfnamefont{S.~J.} \bibnamefont{Pollock}},
  \emph{\bibinfo{title}{Analytic framework for students’ use of mathematics
  in upper-division physics}}, \bibinfo{journal}{Phys. Rev. ST Phys. Educ.
  Res.} \textbf{\bibinfo{volume}{9}}, \bibinfo{pages}{020119}
  (\bibinfo{year}{2013}),
  \urlprefix\url{http://link.aps.org/doi/10.1103/PhysRevSTPER.9.020119}.

\bibitem[{\citenamefont{Pepper et~al.}(2012)\citenamefont{Pepper, Chasteen,
  Pollock, and Perkins}}]{pepper2012math}
\bibinfo{author}{\bibfnamefont{R.~E.} \bibnamefont{Pepper}},
  \bibinfo{author}{\bibfnamefont{S.~V.} \bibnamefont{Chasteen}},
  \bibinfo{author}{\bibfnamefont{S.~J.} \bibnamefont{Pollock}},
  \bibnamefont{and} \bibinfo{author}{\bibfnamefont{K.~K.}
  \bibnamefont{Perkins}}, \emph{\bibinfo{title}{Observations on student
  difficulties with mathematics in upper-division electricity and magnetism}},
  \bibinfo{journal}{Phys. Rev. ST Phys. Educ. Res.}
  \textbf{\bibinfo{volume}{8}}, \bibinfo{pages}{010111} (\bibinfo{year}{2012}),
  \urlprefix\url{http://link.aps.org/doi/10.1103/PhysRevSTPER.8.010111}.

\bibitem[{\citenamefont{Gire and Manogue}(2008)}]{gire2008Qmath}
\bibinfo{author}{\bibfnamefont{E.}~\bibnamefont{Gire}} \bibnamefont{and}
  \bibinfo{author}{\bibfnamefont{C.}~\bibnamefont{Manogue}}, in
  \emph{\bibinfo{booktitle}{Physics Education Research Conference 2008}}
  (\bibinfo{address}{Edmonton, Canada}, \bibinfo{year}{2008}), vol.
  \bibinfo{volume}{1064} of \emph{\bibinfo{series}{PER Conference}}, pp.
  \bibinfo{pages}{115--118}.

\bibitem[{\citenamefont{Zwolak et~al.}(2013)\citenamefont{Zwolak, Kustusch, and
  Manogue}}]{zwolak2013CUErubric}
\bibinfo{author}{\bibfnamefont{J.}~\bibnamefont{Zwolak}},
  \bibinfo{author}{\bibfnamefont{M.~B.} \bibnamefont{Kustusch}},
  \bibnamefont{and} \bibinfo{author}{\bibfnamefont{C.}~\bibnamefont{Manogue}},
  in \emph{\bibinfo{booktitle}{Physics Education Research Conference 2013}}
  (\bibinfo{address}{Portland, OR}, \bibinfo{year}{2013}), PER Conference, pp.
  \bibinfo{pages}{385--388}.

\bibitem[{\citenamefont{{http://www.physics.oregonstate.edu/portfolioswiki/courses}}(2014)}]{paradigms}
\bibinfo{author}{\bibnamefont{{http://www.physics.oregonstate.edu/portfolioswiki/courses}}}
  (\bibinfo{year}{2014}).

\bibitem[{\citenamefont{Wilcox and Pollock}(2013)}]{wilcox2013cue}
\bibinfo{author}{\bibfnamefont{B.}~\bibnamefont{Wilcox}} \bibnamefont{and}
  \bibinfo{author}{\bibfnamefont{S.}~\bibnamefont{Pollock}}, in
  \emph{\bibinfo{booktitle}{Physics Education Research Conference 2013}}
  (\bibinfo{address}{Portland, OR}, \bibinfo{year}{2013}), PER Conference, pp.
  \bibinfo{pages}{365--368}.

\end{thebibliography}
\bibliographystyle{apsper}   

\end{document}